\documentclass[runningheads]{llncs}
\usepackage[T1]{fontenc}
\usepackage{amsmath,amssymb,mathtools}
\usepackage{color}
\usepackage{graphicx} % TODO: try to use EPS format.
\graphicspath{ {./images/} }
\usepackage{subcaption}
\usepackage{float}
\usepackage{hyperref} % hyperlinks

\usepackage{algorithm,algpseudocode} % pseudo-code
\usepackage{listings} % code snippets
\usepackage{multirow} % tables
\begin{document}
\title{Scalability of 3D-DFT by block tensor-matrix multiplication on the JUWELS Cluster}
\author{Nitin Malapally\inst{1,2} \and Viacheslav Bolnykh\inst{1,\dagger} \and Estela Suarez\inst{2,3} \and Paolo Carloni\inst{1,4} \and Thomas Lippert\inst{2,5} \and Davide Mandelli\inst{1,*}}
\authorrunning{N. Malapally et al.}
\titlerunning{Scalability of 3D-DFT by BTMM on the JUWELS Cluster}
\institute{Computational Biomedicine (IAS-5/INM-9), Forschungszentrum J\"ulich, J\"ulich, Germany \and J\"ulich Supercomputing Center (JSC), Institute for Advanced Simulations (IAS), Forschungszentrum J\"ulich, J\"ulich, Germany \and Computer Science Department, University of Bonn, Bonn, Germany \and Molecular Neuroscience and Neuroimaging (INM-11), Forschungszentrum J\"ulich, J\"ulich, Germany \and Frankfurt Institute for Advanced Studies, Goethe Universit\"at Frankfurt, Germany \break \inst{*} Corresponding author: d.mandelli@fz-juelich.de \break \inst{\dagger} Until December 2021}
\maketitle              % typeset the header of the contribution
\begin{abstract} % TODO: cut from 264 to 250 words
The 3D Discrete Fourier Transform (DFT) is a technique used to solve problems in disparate fields. Nowadays, the commonly adopted implementation of the 3D-DFT is derived from the Fast Fourier Transform (FFT) algorithm. However, evidence indicates that the distributed memory 3D-FFT algorithm does not scale well due to its use of all-to-all communication. Here, building on the work of Sedukhin \textit{et al}. [Proceedings of the 30th International Conference on Computers and Their Applications, CATA 2015 pp. 193–200 (01 2015)], we revisit the possibility of improving the scaling of the 3D-DFT by using an alternative approach that uses point-to-point communication, albeit at a higher arithmetic complexity. The new algorithm exploits tensor-matrix multiplications on a volumetrically decomposed domain via three specially adapted variants of Cannon's algorithm. It has here been implemented as a C++ library called S3DFT and tested on the JUWELS Cluster at the J\"ulich Supercomputing Center. Our implementation of the shared memory tensor-matrix multiplication attained 88\% of the theoretical single node peak performance. One variant of the distributed memory tensor-matrix multiplication shows excellent scaling, while the other two show poorer performance, which can be attributed to their intrinsic communication patterns. A comparison of S3DFT with the Intel MKL and FFTW3 libraries indicates that currently iMKL performs best overall, followed in order by FFTW3 and S3DFT. This picture might change with further improvements of the algorithm and/or when running on clusters that use network connections with higher latency, e.g. on cloud platforms.

\end{abstract}
%
%
%
%++++++++++++++++++++++++++++++++++++++++++++
% intro
%++++++++++++++++++++++++++++++++++++++++++++
\section{Introduction}
Many fields of numerical simulation such as astrophysics, plasma physics and molecular dynamics involve computing the pair-wise long-range interactions between the physical system's constituents~\cite{campa2009,yildirim2015,kohnke2020}. Examples are gravitational forces, Van der Waals and electrostatic interactions. This computation is time-consuming and often restricts sizes and time-scales. For example, in atomistic Molecular Dynamics~(MD) simulations of biophysical systems, the size of the system to be simulated can be very large, ranging up to $10^9$~\cite{tuckerman2010} particles, and computing the long-range interactions is responsible for most of the run-time. To limit the computational costs and improve scaling without resorting to truncation schemes, which are known to cause undesirable effects~\cite{arnold2013}, techniques derived from the Ewald summation method are extensively used, which exploit the three dimensional Discrete Fourier Transform (3D-DFT) \textemdash both to ensure convergence of the calculation and to gain speed-up~\cite{toukmaji1996,harvey2009}.

The DFT operation is usually applied using any of the set of algorithms known collectively under the name of Fast Fourier Transform (FFT)~\cite{rockmore2000}. Variants of the Cooley-Tukey FFT algorithm are the most commonly employed. They break the original DFT problem down into a tree of smaller DFT problems, which are solved sometimes recursively and more often non-recursively~\cite{frigo2005}. This results in a drastic reduction of arithmetic complexity from $O(N^2)$ of the na\"ive algorithm, down to $O(Nlog_2N)$. Similarly, the 3D-FFT operation reduces the arithmetic complexity from $O(N^4)$ to $O(N^3log_2N)$. Depending on the size of the problem, this may significantly reduce the run-time of computer applications. However, in the context of distributed memory computers, there is still interest in improving the scaling performance of the 3D-FFT algorithm~\cite{lippert1997,pekurovsky2011,ayala2021}, which is negatively affected by its unavoidable use of all-to-all communications~\cite{jung2016}. Indeed, for massively parallel applications, the algorithm constitutes the bottleneck~\cite{ayala2021}. Specifically, the run-time is dominated by the communication of the algorithm, making up even 80-95\% of it~\cite{pekurovsky2011,ayala2021}. Theoretically, the effect of this high proportion of communication on performance is expected to be stronger at higher node counts. Hence, one could profit from 3D-DFT algorithms that achieve better scalability by making use of alternative communication patterns. 

In this text, we report on the design, implementation and benchmark results of an alternative 3D-DFT algorithm whose performance has been compared with those of modern, state-of-the-art implementations of the FFT algorithm in the context of massively parallel applications.

The paper is organized as follows. Section~\ref{sec:motiv} elaborates on the motivation behind the present work. Section~\ref{sec:theory} introduces the notation used throughout the paper, and reviews key concepts needed to describe our 3D-DFT algorithm. Section~\ref{sec:details} outlines the details of its actual implementation. Section~\ref{sec:analysis} presents a performance analysis of the core functions. Finally, sections~\ref{sec:results} and~\ref{sec:conclusion} discuss the results, and put forward our conclusions.

%++++++++++++++++++++++++++++++++++++++++++++
% motivation
%++++++++++++++++++++++++++++++++++++++++++++
\section{Motivation}
\label{sec:motiv}
Depending on the architecture of the High Performance Computing (HPC) infrastructure and starting with a certain problem size, point-to-point communication scales better than all-to-all communication~\cite{ayala2021}. In order to achieve better scalability by swapping the latter for the former, in 2015, Sedukhin \textit{et al}. studied the scalability of an alternative algorithm which makes use of point-to-point communication to compute the 3D-DFT, albeit at the significantly higher computational complexity of $O(N^4)$, relative to that of $O(N^3\log N)$ of the 3D-FFT algorithm. The authors noted that, for a single node, their implementation of the core computational operation of this algorithm \textemdash the tensor-matrix multiplication \textemdash achieved 20\% of the corresponding peak performance, and concluded with the speculation that a more efficient implementation could outperform the 3D-FFT algorithm for very large node counts~\cite{sedukhin2015}.

Since then, the computational power of CPUs has grown more than the speed of the interconnects between the nodes~\cite{kloeffel2020}, a fact which leans in favour of this alternative approach. However, no other attempts to demonstrate its potential benefits have been published. It is therefore timely to test its scalability on modern machines. Specifically, our work has been conducted on the JUWELS Cluster~\cite{alvarez2021} at the J\"ulich Supercomputing Center, which uses a fat tree network topology \textemdash one of the most commonly adopted network topology nowadays \textemdash whereas the previous study of ref.~\cite{sedukhin2015} made use of the IBM Blue Gene/Q computer, which used a 5D-torus topology.

Here, we have developed an algorithm that we name 3D-DFT by Block Tensor-Matrix Multiplication (BTMM). This is based on specially adapted versions of Cannon's algorithm~\cite{gupta1993,quintin2013,cannon1969}, which, in its original form, is an efficient distributed memory matrix-matrix multiplication algorithm especially well suited for square matrices. We chose this algorithm because of its scalability~\cite{gupta1993} and the simplicity of its implementation. Our adaptations not only make it possible to use tensor operands, but also enable the utilization of the well-known strategy of overlapping communication and computation \textemdash with the help of a custom work-sharing function for OpenMP-based multi-threading \textemdash in an effort to hide the latency of communication. The implementation was designed to make maximum use of the computational resources of the standard compute node of the JUWELS Cluster, on which it was tested and benchmarked, and to simultaneously keep the number of communication events to a minimum.

%++++++++++++++++++++++++++++++++++++++++++++
% theory
%++++++++++++++++++++++++++++++++++++++++++++
\section{Theory}
\label{sec:theory}
%%%%%%%%%%%%%%%%%%%%%%%%%%%%%%%%%%%%%%%%%%%%
\subsection{3D-DFT by Block Tensor-Matrix Multiplication}
\begin{figure}[h!]
    \centering
    \begin{subfigure}{0.4\textwidth}
        \includegraphics[scale=0.3]{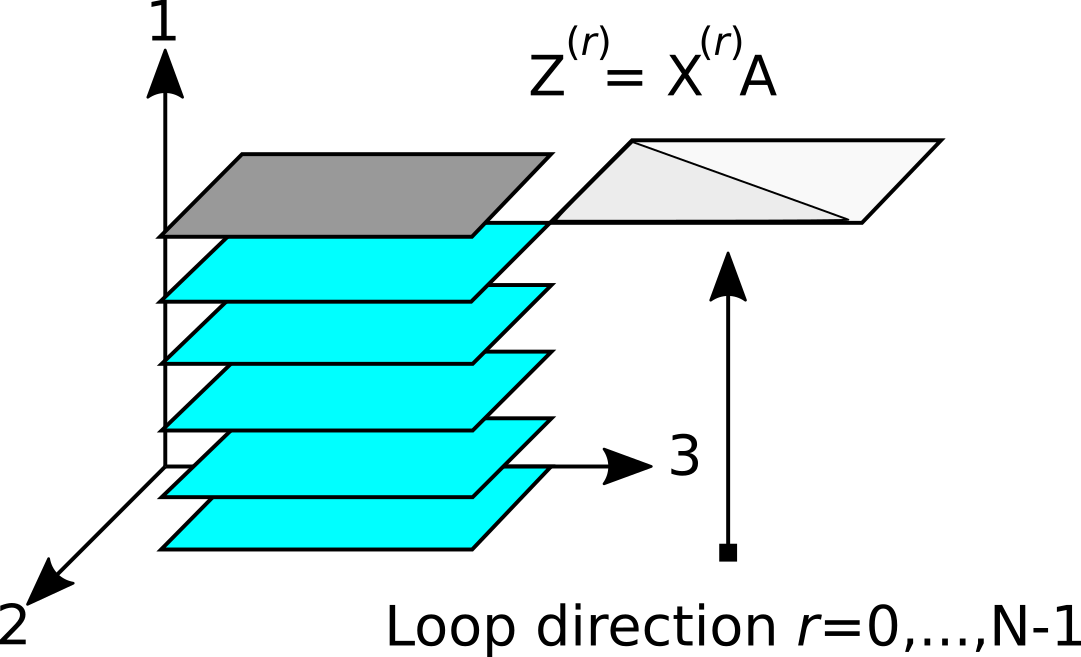}
    \end{subfigure}
    \begin{subfigure}{0.4\textwidth}
        \includegraphics[scale=0.3]{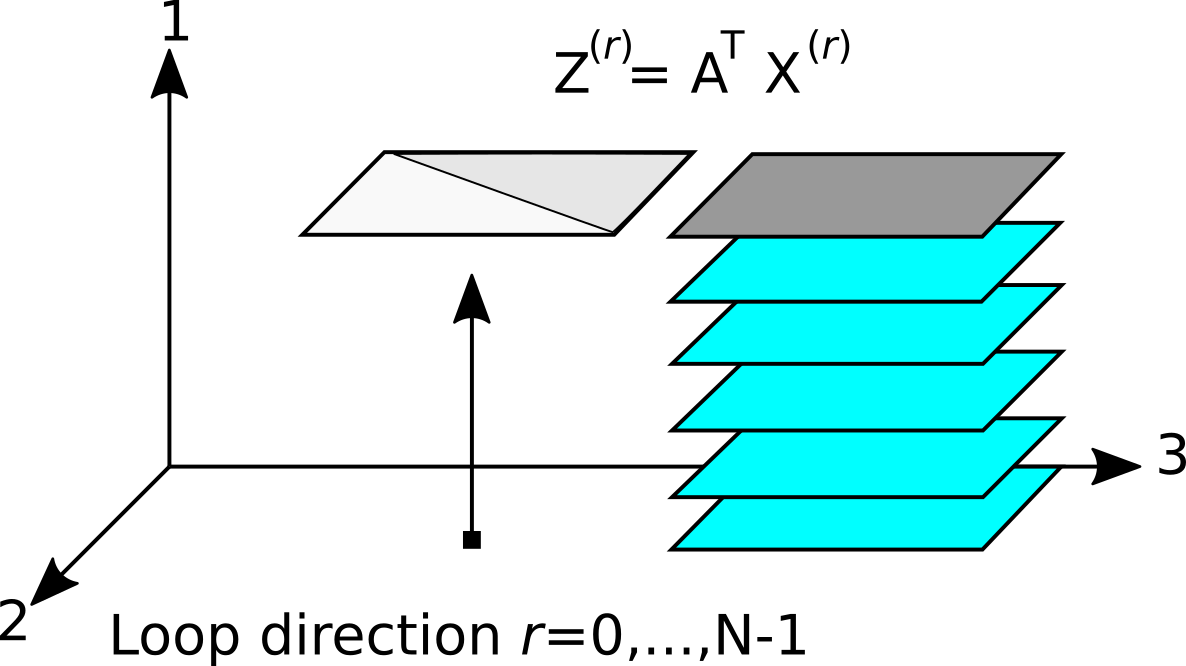}
    \end{subfigure}
    \caption{Visualization of the procedure to compute the tensor-matrix multiplication as a set of independent matrix-matrix multiplications. Left and right panels are for $\rho_R$ and $\rho_L$, respectively.}
    \label{fig:tensor_matrix_mult}
\end{figure}

The DFT for a sequence of 3D operand data $x(l,m,n) \in \bbbc$ can be written as~\cite{sedukhin2015}\footnote{In this work, we only consider the case in which the operand data $x(l,m,n)$ can be arranged as a cube, i.e., $0 \leq l,m,n < N$. To extend the functionality to irregular cuboids, load balancing schemes must be additionally developed, which goes beyond the scope of this work.}
\begin{equation}
    \label{eqn:3d_dft}
    y(i,j,k) = \overset{N-1}{\underset{n=0}{\sum}} \overset{N-1}{\underset{m=0}{\sum}} \overset{N-1}{\underset{l=0}{\sum}} x(l,m,n) \ c(l,i) \ c(m,j) \ c(n,k)
\end{equation}
where the coefficients
\begin{equation*}
    c(n_1,n_2) = \text{exp}(-i \ \frac{2\pi}{N} n_1 n_2) \quad \forall \ 0 \leq n_1,n_2 < N
\end{equation*}
define the DFT matrix $C \in \bbbc^{N \times N}$. Making use of the formalism of Kolda \textit{et al.}~\cite{kolda2009}, the operand data $x(l,m,n)$ can be viewed as an order-3 tensor\footnote{For brevity, henceforth, we shall take "tensor" to mean the order-3 tensor.} $X \in \bbbc^{N \times N \times N}$ existing in a 3D space defined by orthogonal directions $1,2,3$, with the data $x(l,m,n)$ arranged as a cubic mesh. Now, we introduce a right product of $X$ and a matrix $A \in \bbbc^{N \times N}$ in terms of their mode-$3$ product as
\begin{equation}
    \label{eqn:rho_r}
    Z_R = \rho_R(X, A) = X \times_{3} A^T,
\end{equation}
and a left product in terms of their mode-$2$ product as
\begin{equation}
    \label{eqn:rho_l}
    Z_L = \rho_L(X, A) = X \times_{2} A^T,
\end{equation}
where $Z_R, Z_L \in \bbbc^{N \times N \times N}$. These products can be conceived as a set of independent matrix-matrix multiplications as detailed in algorithms \ref{alg:rho_r} and \ref{alg:rho_l}. To visualize this, one can view $X$ as a stack of $N$ matrices $X^{(r)} \in \bbbc^{N \times N}$ for $0 \leq r < N$, which we shall call the slices of the tensor, piled up along any of the 3 orthogonal directions. In practice, to ensure a contiguous memory layout for optimal data-access, we fix the piling direction to direction $1$, as illustrated in figure~\ref{fig:tensor_matrix_mult}.
\begin{algorithm}[t!]
    \begin{algorithmic}[1]
        \For{$r \gets 0$ to $N$}
            \State $Z_R^{(r)} \gets X^{(r)}A$
        \EndFor
    \end{algorithmic}
    \caption{Procedure to perform the tensor-matrix multiplication $\rho_R$.}
    \label{alg:rho_r}
\end{algorithm}
\begin{algorithm}[t!]
    \begin{algorithmic}[1]
        \For{$r \gets 0$ to $N$}
            \State $Z_L^{(r)} \gets A^TX^{(r)}$
        \EndFor
    \end{algorithmic}
    \caption{Procedure to perform the tensor-matrix multiplication $\rho_L$.}
    \label{alg:rho_l}
\end{algorithm}

Using the products defined by equations~\eqref{eqn:rho_r} and~\eqref{eqn:rho_l}, it can be shown that the 3D-DFT equation~\eqref{eqn:3d_dft} can be rewritten in terms of tensor-matrix multiplications as
\begin{equation}
    \label{eqn:3d_dft_tensor_matrix_multiplication}
    Y = \tau(\rho_R(\tau(\rho_L(\rho_R(X, C), C)), C))
\end{equation}
where $X, Y \in \bbbc^{N \times N \times N}$ are the input and output tensors respectively, and $C \in \bbbc^{N \times N}$ is the DFT matrix. The operation $\tau : \bbbc^{N \times N \times N} \rightarrow \bbbc^{N \times N \times N}$ indicates the transposition of the slices along piling direction $2$ of the operand tensor.

In the implementation, the calculation of the transform via equation~\eqref{eqn:3d_dft_tensor_matrix_multiplication} can be performed in three stages~\cite{sedukhin2015}. In the first stage, the tensor-matrix multiplication of the input data with the DFT matrix is carried out as per algorithm~\ref{alg:rho_r},
\begin{equation}
    \label{eqn:stage_1}
    \dot Y = \rho_R(X, C).
\end{equation}
In the second stage, a similar procedure is performed, this time using the output of the first stage, and as per algorithm~\ref{alg:rho_l},
\begin{equation}
    \label{eqn:stage_2}
    \ddot Y = \rho_L(\dot Y, C).
\end{equation}
In the third stage, the piling direction for the tensor-matrix multiplication changes from $1$ to $3$. Consequently, in order to ensure a contiguous memory layout for the subsequent multiplication, a preliminary step must be performed in which $\ddot Y$ is subjected to the transpose operation represented by $\tau$. After this, the final tensor-matrix multiplication is carried out as per algorithm~\ref{alg:rho_r}. At the end, the transpose operation $\tau$ is applied once more to arrange the result in the same spatial layout as the input tensor $X$. Putting everything together, the third and final stage implements the following operations:
\begin{equation}
    \label{eqn:stage_3}
    Y = \tau(\rho_R(\tau(\ddot Y), C)).
\end{equation}
The tensor $Y$ contains the result of the forward DFT operation of equation~\eqref{eqn:3d_dft_tensor_matrix_multiplication}.

%%%%%%%%%%%%%%%%%%%%%%%%%%%%%%%%%%%%%%%%%%%%
\subsection{Adaptation of Cannon's Algorithm for Tensor-Matrix Multiplication}
In this subsection, we provide the designs of the procedures which make use of the basic idea of Cannon's algorithm~\cite{cannon1969} to perform the operations of equations~\eqref{eqn:stage_1},~\eqref{eqn:stage_2},~\eqref{eqn:stage_3}, resulting in three unique distributed memory tensor-matrix multiplication algorithms.

\begin{figure}[t!]
    \begin{center}
        \includegraphics[scale=0.3]{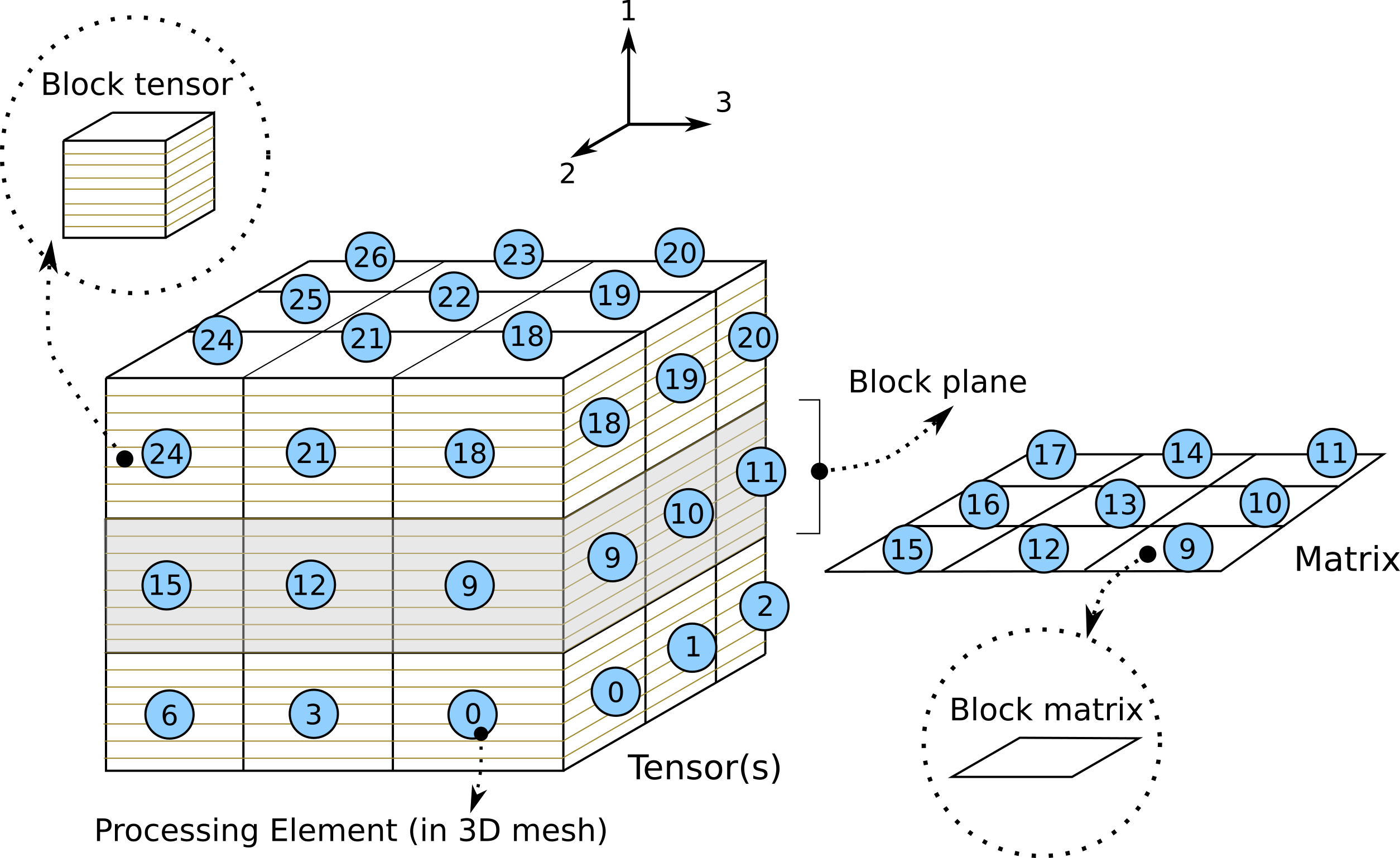}
    \end{center}
    \caption{Volumetric decomposition: a tensor and a matrix are broken down into $p^3$ and $p^2$ blocks, respectively (here, $p=3$). Each block is locally allocated for by a corresponding PE, as indicated in the circles.}
    \label{fig:volumetric_decomposition}
\end{figure}

In order to enable the use of as many Processing Elements (PEs) as possible, the tensors $X$ and $Y$ from equation~\eqref{eqn:3d_dft_tensor_matrix_multiplication} are subjected to the volumetric domain decomposition~\cite{sedukhin2015,ayala2021,jung2016}, resulting in a cubic mesh comprising $p^3$ PEs (see  figure~\ref{fig:volumetric_decomposition}). Each PE$(i,j,k)$ is accessible via indices $\ 0 \leq i,j,k < p$, and has locally allocated operand and result block tensors $X_{i,j,k}^{(b)}, Y_{i,j,k}^{(b)} \in \bbbc^{b \times b \times b}$, and an operand block matrix $C_{j,k}^{(b)} \in \bbbc^{b\times b}$, which is obtained by matrix decomposition. Here, the block size is given by $b=N/p$.

We will start by focusing on the procedure of the multiplication involved in the first stage, given by equation~\eqref{eqn:stage_1}. From the description of $\rho_R$ in algorithm~\ref{alg:rho_r} we recall that the slices of $X$ along piling direction 1 are to be multiplied with $C$ (see also figure~\ref{fig:tensor_matrix_mult}). Analogously, here, the block \textit{tensors} in each plane orthogonal to piling direction 1 are to be multiplied with $C$. The multiplication of each such block plane with $C$ can be executed independently \textemdash and hence \textemdash parallely. It is for this multiplication that we can utilize a scheme almost identical to that of the original Cannon's algorithm, with the essential deviation that the corresponding operand and result matrices be replaced by operand and result tensors, respectively. The adapted algorithm is composed of an alignment and a computation phase, with a total of $p$ communication events. The communication events of the two phases involve $2p^2(p - 1)$ and $2p^3$ parallel communication calls respectively\footnote{Conceptually, parallel communication calls are executed simultaneously by all PEs and thus the duration of a single communication event is decided by the most time-consuming communication call.}.
\begin{algorithm}[t!]
    \begin{algorithmic}[1]
        \Statex \(\triangleright\) Alignment phase:
        \If{$j \neq 0$}
            \State send $X_{i,j,k}^{(b)} : \quad \text{PE}(i,j,k) \rightarrow \text{PE}(i,j,(k+p-j) \pmod p)$
        \EndIf
        \If{$k \neq 0$}
            \State send $\hphantom{X_{i,j,k}^{(b)}}\mathllap{C_{j,k}^{(b)}} : \quad \text{PE}(i,j,k) \rightarrow \text{PE}(i,(j+p-k) \pmod p,k)$
        \EndIf
        \Statex \(\triangleright\) Computation phase:
        \For{$r \gets 0$ to $p$}
            \State $Y_{i,j,k}^{(b)} \gets Y_{i,j,k}^{(b)} + X_{i,j,k}^{(b)} * C_{j,k}^{(b)}$
            \If{$r < p-1$}
                \State send $X_{i,j,k}^{(b)} : \quad \text{PE}(i,j,k) \rightarrow \text{PE}(i,j,(k+p-1) \pmod p)$
                \State send $\hphantom{X_{i,j,k}^{(b)} }\mathllap{C_{j,k}^{(b)}} : \quad \text{PE}(i,j,k) \rightarrow \text{PE}(i,(j+p-1) \pmod p,k)$
            \EndIf
        \EndFor
    \end{algorithmic}
    \caption{Procedure of the adapted Cannon's algorithm for the function $Y = \rho_R(X, C)$. Variables $X_{i,j,k}$ and $Y_{i,j,k}$ are the locally allocated operand and result block tensors, and $C_{j,k}$ the operand block matrix.}
    \label{alg:adapted_cannonalgo_rho_r}
\end{algorithm}
\begin{algorithm}[t!]
    \begin{algorithmic}[1]
        \Statex \(\triangleright\) Alignment phase:
        \State send $C_{j,k}^{(b)} : \quad \text{PE}(i,j,k) \rightarrow \text{PE}(i,k,(j+p-k) \pmod p)$
        \If{$k \neq 0$}
            \State send $X_{i,j,k}^{(b)} : \quad \text{PE}(i,j,k) \rightarrow \text{PE}(i,(j+p-k) \pmod p,k)$
        \EndIf
        \Statex \(\triangleright\) Computation phase:
        \For{$r \gets 0$ to $p$}
            \State $Y_{i,j,k}^{(b)} \gets Y_{i,j,k}^{(b)} + C_{j,k}^{(b) T} * X_{i,j,k}^{(b)}$
            \If{$r < p-1$}
                \State send $\hphantom{X_{i,j,k}^{(b)}}\mathllap{C_{j,k}^{(b)}} : \quad \text{PE}(i,j,k) \rightarrow \text{PE}(i,j,(k+p-1) \pmod p)$
                \State send $X_{i,j,k}^{(b)} : \quad \text{PE}(i,j,k) \rightarrow \text{PE}(i,(j+p-1) \pmod p,k)$
            \EndIf
        \EndFor
    \end{algorithmic}
    \caption{Procedure of the adapted Cannon's algorithm for the function $Y = \rho_L(X, C)$. Variables $X_{i,j,k}$ and $Y_{i,j,k}$ are the locally allocated operand and result block tensors, and $C_{j,k}$ the operand block matrix.}
    \label{alg:adapted_cannonalgo_rho_l}
\end{algorithm}
\begin{algorithm}[t!]
    \begin{algorithmic}[1]
        \Statex \(\triangleright\) First local transposition:
        \State $X_{i,j,k}^{(b)} \gets \tau(X_{i,j,k}^{(b)})$
        \Statex \(\triangleright\) Alignment phase:
        \If{$k \neq 0$}
            \State Send $X_{i,j,k}^{(b)} : \quad \text{PE}(i,j,k) \rightarrow \text{PE}((i+p-j) \pmod p,j,k)$
        \EndIf
        \State Send $C_{j,k}^{(b)} : \quad \text{PE}(i,j,k) \rightarrow \text{PE}(k,(j+p-k) \pmod p, i)$
        \Statex \(\triangleright\) Computation phase:
        \For{$r \gets 0$ to $p$}
            \State $Y_{i,j,k}^{(b)} \gets Y_{i,j,k}^{(b)} + X_{i,j,k}^{(b)} * C_{j,k}^{(b)}$
            \If{$r < p-1$}
                \State send $X_{i,j,k}^{(b)} : \quad \text{PE}(i,j,k) \rightarrow \text{PE}((i+p-1) \pmod p,j,k)$
                \State send $\hphantom{X_{i,j,k}^{(b)}}\mathllap{C_{j,k}^{(b)}} : \quad \text{PE}(i,j,k) \rightarrow \text{PE}(i,(j+p-1) \pmod p,k)$
            \EndIf
        \EndFor
        \Statex \(\triangleright\) Final local transposition:
        \State $Y_{i,j,k}^{(b)} \gets \tau(Y_{i,j,k}^{(b)})$
    \end{algorithmic}
    \caption{Procedure of the adapted Cannon's algorithm for the function $Y = \tau(\rho_L(\tau(X), C))$. Variables $X_{i,j,k}$ and $Y_{i,j,k}$ are the locally allocated operand and result block tensors, and $C_{j,k}$ the operand block matrix.}
    \label{alg:adapted_cannonalgo_transposed_rho_r}
\end{algorithm}

The procedure is as outlined in algorithm~\ref{alg:adapted_cannonalgo_rho_r}. The local update in step 8 can be performed slice-wise as indicated by algorithms~\ref{alg:rho_r} and~\ref{alg:rho_l}, either sequentially, or, when additional computational resources are available to each PE, using a separate mode of parallelism, giving rise to multi-level parallelism. Additionally, the communication event (from steps 10 and 11) and the local update can also be executed in parallel. At the end of the computation phase, the block tensor $Y^{(b)}_{i,j,k}$ located in each PE contains the block result of the desired product.

The procedure for the multiplication of the second stage, given by equation~\eqref{eqn:stage_2}, is outlined in algorithm~\ref{alg:adapted_cannonalgo_rho_l}. It is very similar to that of the first stage, albeit with minor changes in the pattern of communication. The procedure for the third stage needs to incorporate the transpose operations represented by $\tau$ in equation~\eqref{eqn:stage_3}. Although these data transpositions are unavoidable and involve communication, we can eliminate the majority of this overhead by \textit{transposing the mesh of the PEs instead of the data itself}. All the same, a local (i.e., not involving communication) data transpose function is required to perform the transposition of the operand and result block tensors. The procedure of the adapted Cannon's algorithm for the third stage is provided in algorithm~\ref{alg:adapted_cannonalgo_transposed_rho_r}.

%++++++++++++++++++++++++++++++++++++++++++++
% details of the implementation
%++++++++++++++++++++++++++++++++++++++++++++
\section{Details of the Implementation and Computer}
\label{sec:details}
%%%%%%%%%%%%%%%%%%%%%%%%%%%%%%%%%%%%%%%%%%%%
The 3D-DFT by BTMM algorithm was implemented as a C++ library named S3DFT, which offers a distributed memory Application Programming Interface (API). The library was built and run using the Intel compiler and the Intel MPI library, which are part of the Intel OneAPI v2021.4.0 toolkit suite. S3DFT is open-source software available for use under the GNU Lesser General Public License v3 (LGPL)~\cite{s3dft}.

Benchmarks and tests have been executed on the JUWELS Cluster~\cite{alvarez2021}. The standard compute node has two cache-coherent NUMA domains. The salient specifications are listed in table~ \ref{tab:specifications_standard_compute_node}, and the peak bandwidths as obtained with the Intel$^{\circledR}$ Advisor tool~\cite{intel2022}, in table~\ref{tab:xeon_8168_intel_advisor_bandwidths}. These numbers were used as reference values while analysing the performance of the core functions of the implementation.
\begin{table}[b!]
    \centering
    \caption{Specifications of the standard compute node of the JUWELS cluster}
    \label{tab:specifications_standard_compute_node}
    \begin{tabular}{| c | c |}
        \hline
        %Component/Detail & Specification \\
        %\hline
        Processor & Intel$^{\circledR}$ Xeon$^{\circledR}$ Platinum 8168 (Skylake) \\
        \hline
        CPU count & 2 (sockets)\\
        \hline
        Core count & 48 cores (24 cores per CPU) \\
        \hline
        SMT/HT & Available, 96 threads (48 threads per CPU) \\
        \hline
        Clock frequency & [1.2 - 3.7 GHz], base @ 2.7 GHz \\
        \hline
        Cache & L1 - 32 kB, L2 - 1 MB, L3 - 33 MB \\
        \hline
        DRAM & 96 GB DDR4 @ 2666 MHz \\
        \hline
    \end{tabular}
\end{table}
\begin{table}[t!]
    \centering
    \caption{Cache and memory bandwidths according to Intel$^{\circledR}$ Advisor.}
    \label{tab:xeon_8168_intel_advisor_bandwidths}
    \begin{tabular}{| c | c | c |}
        \hline
        \multirow{2}{*}{} & \multicolumn{2}{c|}{Bandwidth} \\
        \cline{2-3}
        & 1x NUMA & 2x NUMA \\
        \hline
        L1 & 11.6 TB/s & 23.2 TB/s \\
        \hline
        L2 & 5.5 TB/s & 11.0 TB/s \\
        \hline
        L3 & 649 GB/s & 1299 GB/s \\
        \hline
        DRAM & 115 GB/s & 230 GB/s \\
        \hline
    \end{tabular}
\end{table}

Since Intel did not explicitly provide information on peak performance in terms of FLOP/s at the time of writing of this article~\cite{intel2021}, the peak performance of the compute node had to be estimated using the published specifications. The base frequency of the Intel$^{\circledR}$ Xeon$^{\circledR}$ Platinum 8168 processor is 2.7~GHz~\cite{intel2020}. However, when all cores are active and the use of AVX-512 instructions is maximized, the clock frequency drops to 2.5~GHz~\cite{intel2020}. The processor is equipped with 2 AVX-512 Fused Multiply-Add (FMA) units per core, which yields a theoretical peak performance of 3840~GFLOP/s. Corroborating this estimation, Intel$^{\circledR}$ Advisor's roof-line chart includes information about the double-precision FMA peak performance~\cite{intel2022}, which in this case is 3812~GFLOP/s. Henceforth, we refer to this value when we speak about the peak performance of the node.

S3DFT uses both shared and distributed memory parallelism to minimize the number of communication events, while simultaneously maximising the utilization of hardware resources. Keeping this in mind, an OpenMP/MPI hybrid approach was selected such that the implementation can use OpenMP-based shared memory parallelism across multiple NUMA domains within a single compute node, if and when such is the case.

For the matrix-matrix multiplication within the shared memory tensor-matrix multiplication, as represented by algorithms~\ref{alg:rho_r} and~\ref{alg:rho_l}, we used the CBLAS implementation provided by the Intel$^{\circledR}$ Math Kernel Library (MKL) v2021.4.0. We see from the strong scaling of the shared memory tensor-matrix multiplication (right panel of figure~\ref{fig:scaling_sm_tmm}) that the difference in performance when all 48 cores are used and when 47 cores are used is small, with the performance decreasing slightly from 88\% to 85\% of the single node peak performance. Hence, we decided to \textit{dedicate one thread to communication}, which we implemented by means of a custom work-sharing function for OpenMP parallel regions. This was done in an effort to hide the latency of communication, by overlapping the communication event and the local update in the implementations of algorithms \ref{alg:adapted_cannonalgo_rho_r}, \ref{alg:adapted_cannonalgo_rho_l} and \ref{alg:adapted_cannonalgo_transposed_rho_r}.

%++++++++++++++++++++++++++++++++++++++++++++
% micro-benchmarking analysis
%++++++++++++++++++++++++++++++++++++++++++++
\section{Micro-benchmarking Analysis}
\label{sec:analysis}
In this section, we present the performance analysis of the core functions of the S3DFT implementation. The open-source library TiXL, which is available under the LGPL v3, was used for this purpose~\cite{tixl}.

The micro-benchmark programs consisted of (i) an initialization phase, in which operand/result data were allocated afresh, and each thread (excepting the communication thread) accessed the first word of each memory page of its associated data \textemdash thereby ruling out the possibility of measuring page-faults, (ii) an experiment phase, in which the function of interest was run, and (iii) a clean-up phase in which all data were freed. Each benchmark test was concluded by performing 20 warm-up and 100 timed runs. Only the latter were used to measure durations in the experiment phase. The result was calculated as the arithmetic mean of these measurements.

%%%%%%%%%%%%%%%%%%%%%%%%%%%%%%%%%%%%%%%%%%%%
\subsection{Transpose Function}
\label{subsec:transpose_function}
\begin{figure}[t!]
\centering
    \begin{subfigure}{0.49\textwidth}
        \includegraphics[scale=0.16]{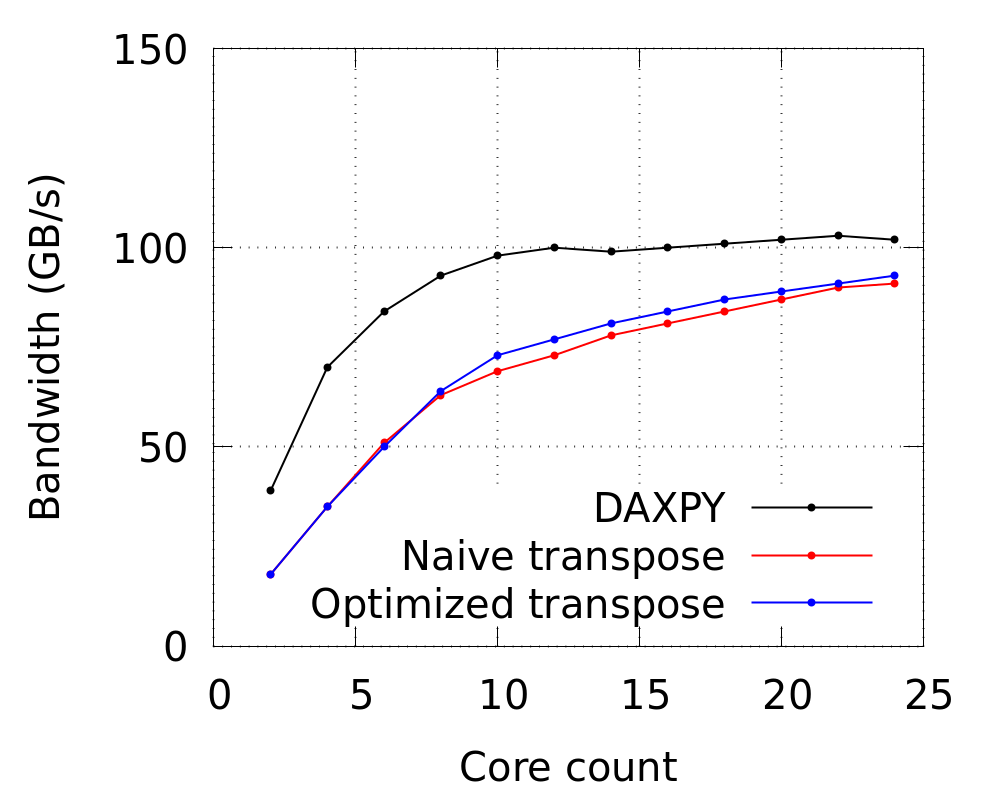}
    \end{subfigure}
    \begin{subfigure}{0.49\textwidth}
        \includegraphics[scale=0.16]{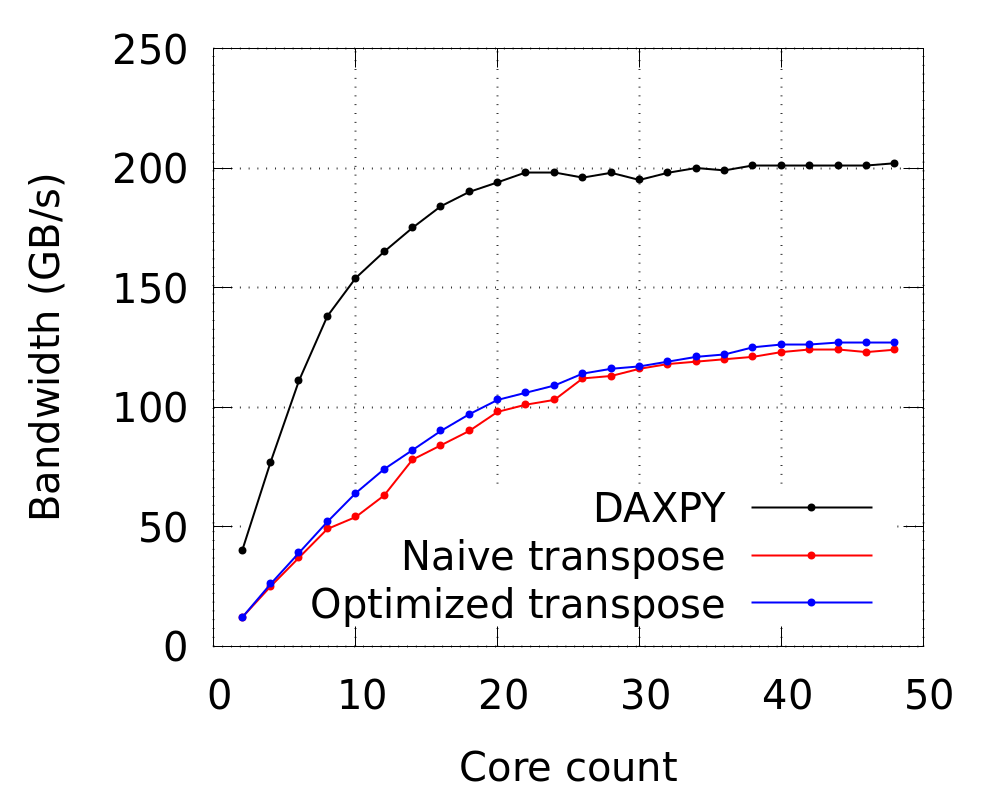}
    \end{subfigure}
    \caption{Comparison of the effective bandwidths achieved by the DAXPY kernel and the na\"ive and optimized versions of the transpose function. Left and right panels report results obtained in single and dual NUMA domains configurations, respectively.}
    \label{fig:juwels_bandwidth_gbs}
\end{figure}

Here, we take a closer look at the performance of our implementation of the transpose function, $\tau$ of equation~\eqref{eqn:stage_3}. The transpose function can be viewed as a streaming function because in a perfect implementation it would closely resemble a copy operation. Thus, one way to asses the performance of our implementation is to compare it to that of a suitably similar streaming function having an excellent memory bandwidth utilization. As a reference, we decided to use the DAXPY loop, by measuring the performance of the operation $Y = Y + aX$ on the target computer system, where $Y,X \in \bbbr^n$ are double-precision arrays and $a \in \bbbr$ is a double-precision scalar. We chose $n=950,700$ for the dual and single NUMA configurations, respectively, which are sizes at which the performance of the loop was found to saturate. The black lines in figure \ref{fig:juwels_bandwidth_gbs} illustrate the increase of the effective bandwidth\footnote{The effective bandwidth is calculated using the run-time duration and the data-traffic estimation.} of this reference loop as a function of the number of cores, for the single as well as dual NUMA configurations. The corresponding recorded peak performances are 103 GB/s and 202 GB/s, respectively. We used these values as reference to measure the efficiency of our implementation of the transpose function.

The performance of the na\"ive implementation of the transpose function is shown by the red curves in figure~\ref{fig:juwels_bandwidth_gbs} for sizes $N=1300,700$ for the single and dual NUMA configurations, respectively. Building on it, we improved the cache utilization by applying loop-blocking with the help of an intermediate array so small as to fit into the cache. The optimal blocking size was experimentally found to be 16~kiB. Upon optimization, only a small improvement in performance could be observed, as shown by the blue curves in figure~\ref{fig:juwels_bandwidth_gbs}. Indeed, we found the performance of the n\"aive implementation to be quite high, which we attributed to the size of the processor's L3-cache, by virtue of which good cache-line reuse can be achieved even for relatively large matrix sizes.

In the single NUMA domain configuration, the optimized transpose function attained a peak efficiency of 91\% as compared to that of 89\% of the naive function. However, we note that the efficiencies drop to 63\% and 61\% respectively, when the dual NUMA configuration is applied. This can be attributed to unavoidable non-local memory accesses arising from the fact that the functions make use of a different multithreading work-sharing plan as compared to the other core functions in the implementation. Although an adaptation of the algorithm to minimize these non-local memory accesses is conceivable, the expected performance gain did not justify its design and implementation within the scope of this study.

%%%%%%%%%%%%%%%%%%%%%%%%%%%%%%%%%%%%%%%%%%%%
\subsection{Shared Memory Tensor-Matrix Multiplication}
This function performs the local update operation of the distributed memory tensor-matrix multiplication as indicated by step~8, step~6 and step~7 of algorithms \ref{alg:adapted_cannonalgo_rho_r}, \ref{alg:adapted_cannonalgo_rho_l} and \ref{alg:adapted_cannonalgo_transposed_rho_r}, respectively.
\begin{figure}[t!]
\centering
    \begin{subfigure}{0.4\textwidth}
        \centering
        \includegraphics[scale=0.17]{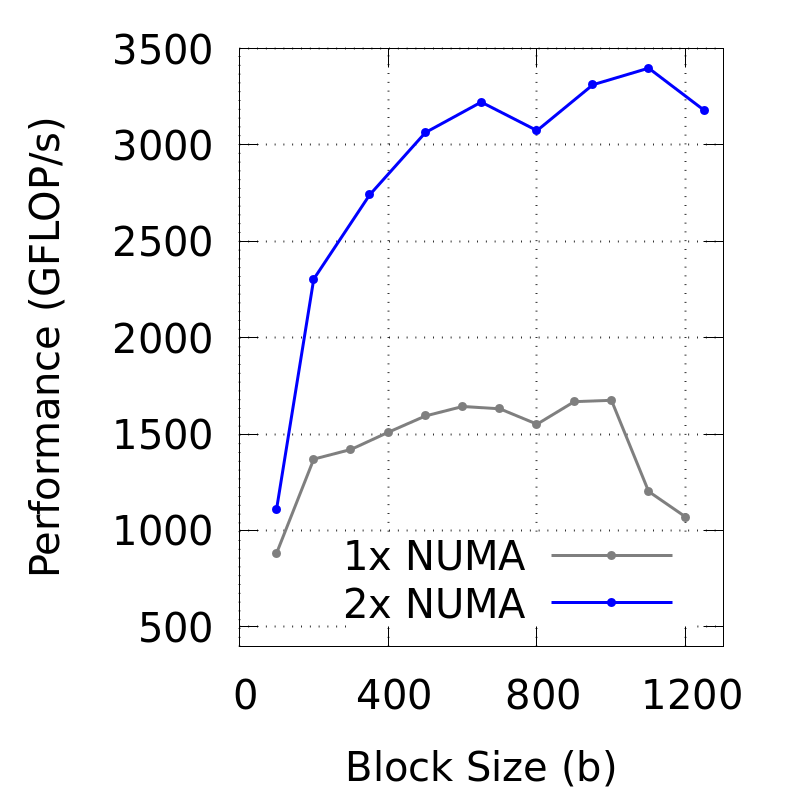}
    \end{subfigure}
    \begin{subfigure}{0.4\textwidth}
        \centering
        \includegraphics[scale=0.17]{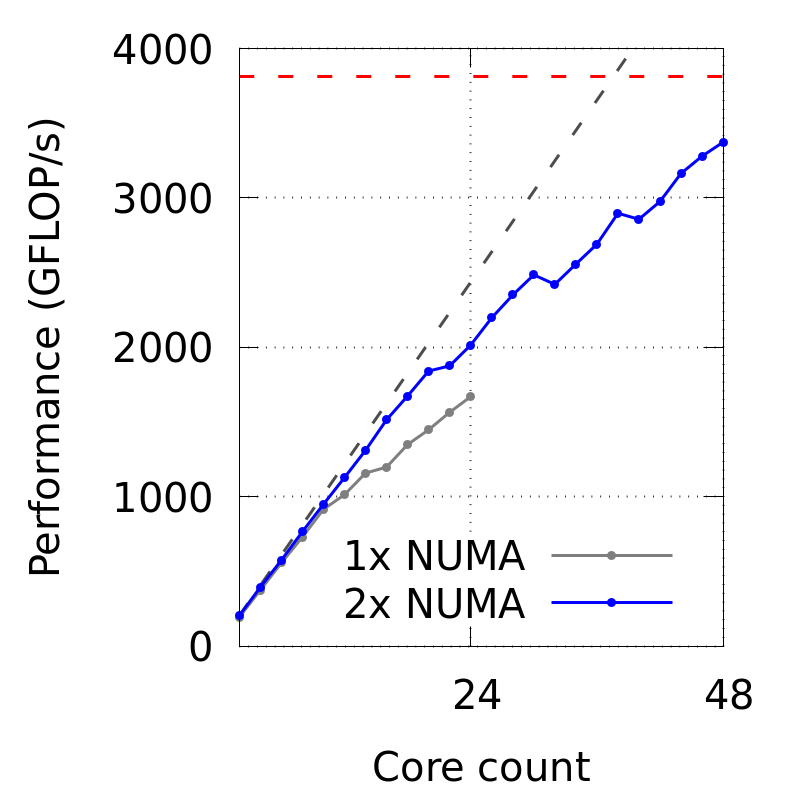}
    \end{subfigure}
    \caption{The left panel shows the performance of the shared memory tensor-matrix multiplication as a function of block size and the right panel, its strong scaling behaviour. Grey and blue curves are for single (24 cores) and dual (48 cores) NUMA confgurations, respectively. The corresponding problem sizes are $N=900,1100$. The red line in the right panel indicates the peak performance of the single node. The grey dashed line indicates the ideal linear scaling.}
    \label{fig:scaling_sm_tmm}
\end{figure}

We began the analysis by running problem scaling tests to identify the problem sizes at which peak performances of the function can be expected. Next, we conducted strong scaling tests for these problem sizes. The results  are reported in figure \ref{fig:scaling_sm_tmm}. We observed a peak performance of $1670$~GFLOP/s at $N \sim 900$ and of $3371$~GFLOP/s at $N \sim 1100$ for the single and dual NUMA configuration, respectively, corresponding to 88\% of the peak performance of the single node. As shown in the right panel of figure~\ref{fig:scaling_sm_tmm}, the function scales well.

To estimate the corresponding effective bandwidth, let us first model the traffic and computation requirements of algorithm~\ref{alg:rho_r}. For a tensor of side $N$, a computer could perform $8N^4$ floating point operations\footnote{Each complex number addition and multiplication involves at least 2 and 6 FLOPs respectively.} after $2N^3(N + 1)$ transfers. Assuming double-precision, we have the code balance given by $B_c = \frac{4(N + 1)}{N} \approx 4$ B/FLOP. Using the roof-line model, we can calculate the effective bandwidth as $b_s = B_c P$, where $P$ is the attained performance~\cite[p.~66]{hager2010}. Following this, we can estimate peak effective bandwidths $b_s = 6.7\text{~TB/s}$ and $b_s = 13.5\text{~TB/s}$ for the single and dual NUMA configurations, respectively, which are greater than the corresponding L2-cache bandwidths as listed in table~\ref{tab:xeon_8168_intel_advisor_bandwidths}. This can be taken to conclude that the function makes excellent use of caching.

%%%%%%%%%%%%%%%%%%%%%%%%%%%%%%%%%%%%%%%%%%%%
\subsection{Distributed Memory Tensor-Matrix Multiplication}
\label{subsec:dm_tmm}
In this analysis, we consider the implementations of algorithms~\ref{alg:adapted_cannonalgo_rho_r},~\ref{alg:adapted_cannonalgo_rho_l} and~\ref{alg:adapted_cannonalgo_transposed_rho_r}.

\begin{figure}[h!]
\centering
    \begin{subfigure}{0.4\textwidth}
        \centering
        \includegraphics[scale=0.17]{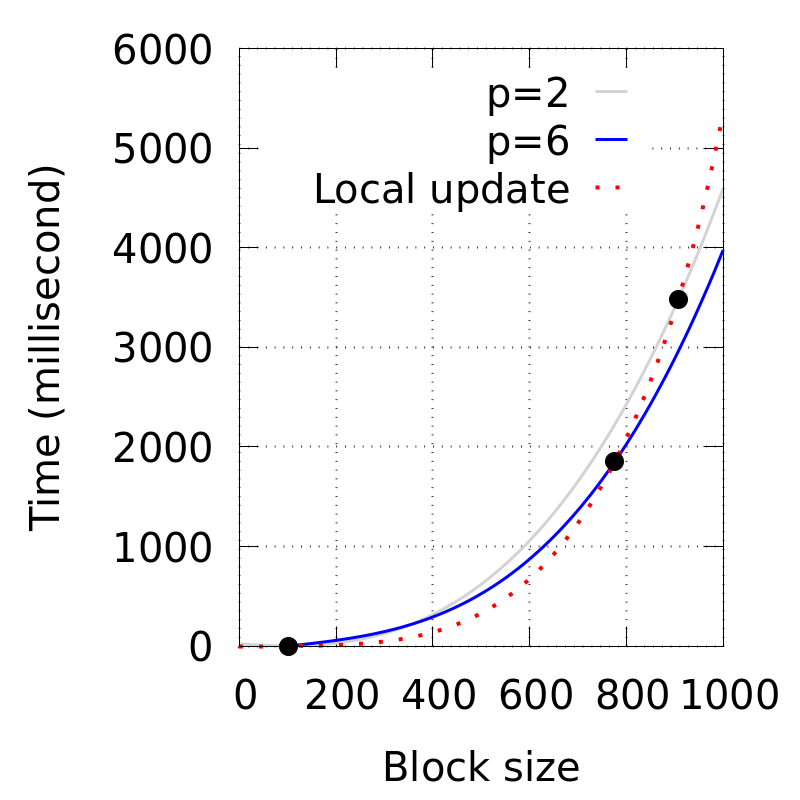}
    \end{subfigure}
    \begin{subfigure}{0.4\textwidth}
        \centering
        \includegraphics[scale=0.17]{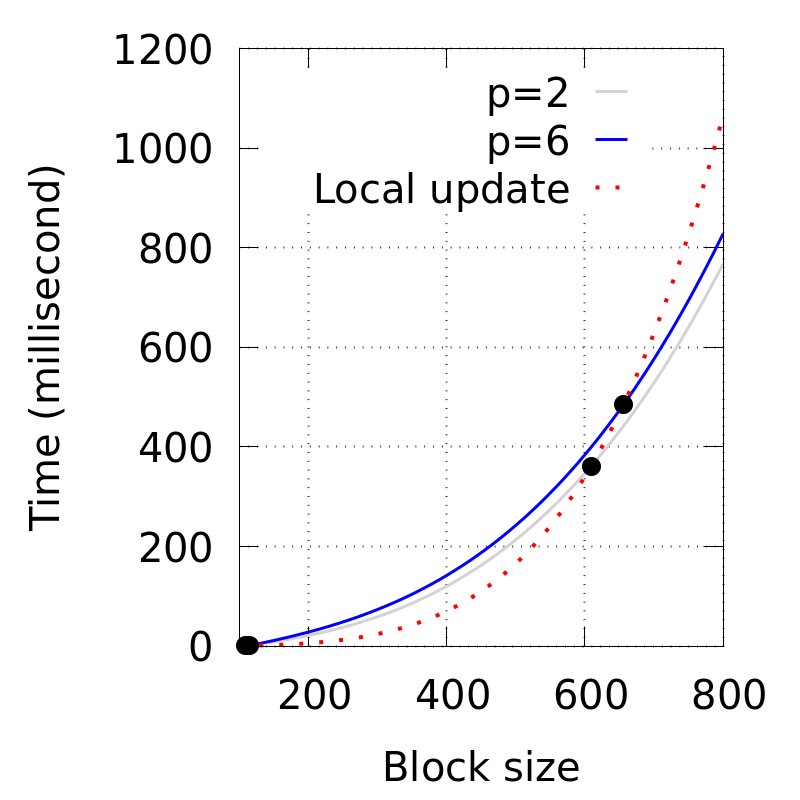}
    \end{subfigure}
    \caption{Fitted curves showing the duration of the communication event (continuous lines, different node counts) and that of the local update (dotted line) as functions of block size. The black dots indicate the intersections at which perfect overlapping can be expected. Left and right panels report results obtained in 2 MPI tasks/node and 1 MPI task/node configurations, respectively.}
    \label{fig:comm_dm_tmm}
\end{figure}

First, we identify the configurations of block size and node count for which the highest efficiency of the algorithm can be reached. For this purpose, we designed micro-benchmark programs which exactly imitate step $8$ (local update), and steps $10$ and $11$ (communication event) in algorithm~\ref{alg:adapted_cannonalgo_rho_r}, and ran them for block sizes in the interval $[100,1300]$. We then fitted the results with cubic polynomials to model the run-time duration of the communication event for various node counts (continuous curves in figure~\ref{fig:comm_dm_tmm}), and a quartic polynomial for that of the local update (dotted curves). The points of intersection identify those configurations at which theoretical peak efficiencies can be expected because they represent the conditions under which the algorithms do not incur communication overhead. Using this technique, the optimal block sizes were found to be $b \sim 100, \sim 800$ and $b \sim 110, \sim 630$ for the 2 MPI tasks/node (single NUMA) and 1 MPI task/node (dual NUMA) configurations, respectively.
\begin{figure}[h!]
\centering
    \begin{subfigure}{0.3\textwidth}
        \includegraphics[scale=0.173]{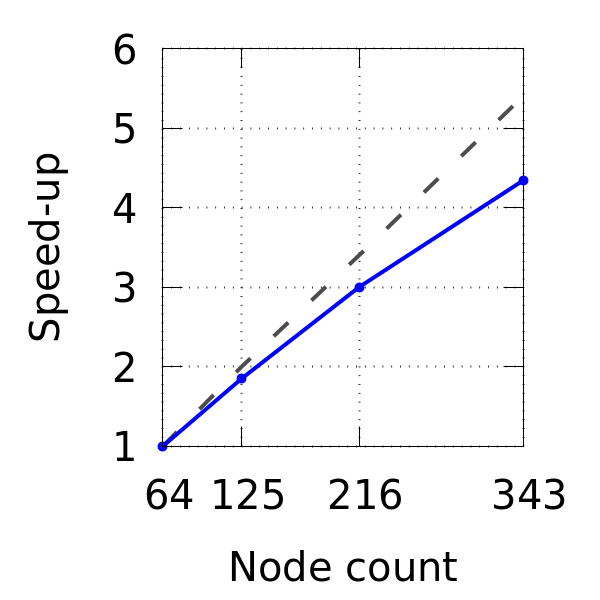}
    \end{subfigure}
    \begin{subfigure}{0.3\textwidth}
        \includegraphics[scale=0.173]{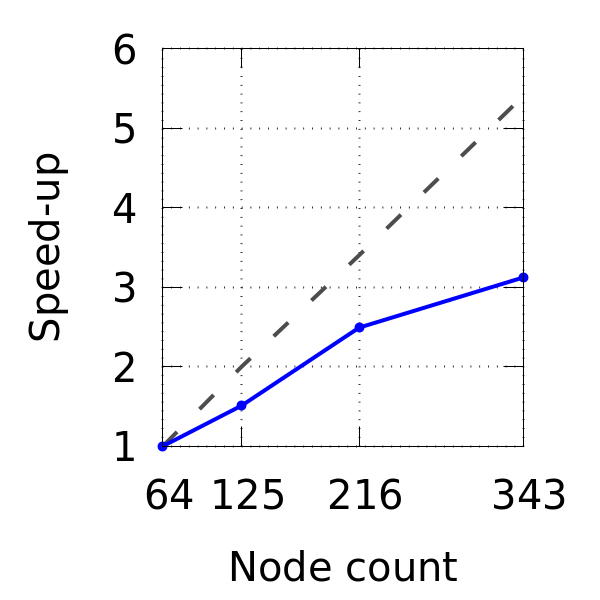}
    \end{subfigure}
    \begin{subfigure}{0.3\textwidth}
        \includegraphics[scale=0.173]{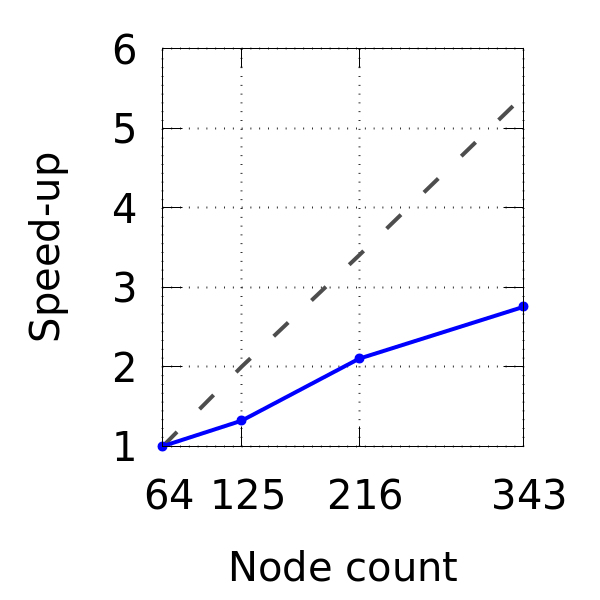}
    \end{subfigure}
    \caption{The panels from left to right show the strong scaling performance of the implementations of algorithm~\ref{alg:adapted_cannonalgo_rho_r}, algorithm~\ref{alg:adapted_cannonalgo_rho_l} and~algorithm~\ref{alg:adapted_cannonalgo_transposed_rho_r}, respectively, at problem size $N=4200$, in the 1 MPI task/node configuration. The grey dashed lines indicate the ideal linear scaling.}
    \label{fig:ss_dm_tmm}
\end{figure}

Next, we conducted strong scaling tests for all three variants in the 1 MPI task/node configuration for the problem size $N=4200$ and $p=4,5,6,7$, corresponding to block sizes $b=1050,840,700,600$, at which the latency of communication is expected to be hidden by the overlapping, as shown in the right panel of figure~\ref{fig:comm_dm_tmm}. The results are reported in figure~\ref{fig:ss_dm_tmm}, showing parallel efficiencies\footnote{Here, the parallel efficiency has been evaluated relative to $p=4$ case i.e. with 64 nodes, which is the minimum number of nodes we could use for the given problem size due to memory limitations.} in the ranges of $81\% - 95\%$, $58\% - 77\%$ and $51\% - 68\%$ for algorithms~\ref{alg:adapted_cannonalgo_rho_r},~\ref{alg:adapted_cannonalgo_rho_l} and~\ref{alg:adapted_cannonalgo_transposed_rho_r}, respectively. Further investigations indicated that the poor scaling behaviour of algorithms~\ref{alg:adapted_cannonalgo_rho_l} and~\ref{alg:adapted_cannonalgo_transposed_rho_r} can be attributed to their communication patterns. More precisely, we found that although the communication always takes place between neighbours that are equidistant along each direction in the 3D mesh of PEs, the latency of communication varies strongly depending on the direction along which these neighbours are identified. This is because they are not equidistant in the topology of the hardware allocation. Specifically, our experiments indicated that communication was fastest for neighbouring PEs along direction 3 and slowest for those along direction 1.

In an effort to improve the scalability of algorithms~\ref{alg:adapted_cannonalgo_rho_l} and~\ref{alg:adapted_cannonalgo_transposed_rho_r}, we designed and tested variants in which the majority of the communication occurs between neighbours along direction 3. This is made possible by replacing the transposition of the PE-mesh by global data transpositions, with an additional communication event. Although this improved the strong scaling parallel efficiency of stage 2 and 3 to $\sim80$\% and $\sim95$\%, respectively, the overall performance was found to be similar, owing to a worsening of the performance at lower node counts. We eventually decided to retain the original algorithms for the final implementation. 

On a different note, we observe that the performance of the local update reduces with reducing block size (see the left panel of figure~\ref{fig:scaling_sm_tmm}), which warns us that the performance of the above-mentioned algorithms could be strongly reduced when the node count is increased while keeping the problem size constant.

%++++++++++++++++++++++++++++++++++++++++++++
% results and conclusions
%++++++++++++++++++++++++++++++++++++++++++++
\section{Results}
\label{sec:results}
We have tested S3DFT against two competitive 3D-FFT implementations: the FFTW3 v3.3.10 and Intel$^{\circledR}$ MKL v2021.4.0\footnote{We made use of the convenient FFTW3 wrapper interface provided by Intel$^{\circledR}$ MKL, which makes use of its implementation of cluster FFT functions.}. The benchmarking procedure is identical to that outlined in section~\ref{sec:analysis}, with the exception that 70 warm-up runs and 50 timed experiments were conducted. In the plots, we report the range between the minimum value recorded and the arithmetic mean. In the programs which recorded the performance of the cluster-based FFTW3/iMKL libraries, multithreading was initialized as per the manual~\cite{fftw3_manual}. The FFTW-plan~\cite{frigo2005} was created in the initialization phase of the program using the flag \lstinline{FFTW_MEASURE}. To be able to benchmark under reproducible conditions, contiguous node allocation was requested. Within a single compute node, a thread-placement policy of 1 thread/core was applied. Further, each thread was pinned to avoid being migrated by the operating system during run-time. This was done since allowing the free migration of threads across NUMA domains within the node would have caused poor performance owing to excessive non-local data accesses.

Initial testing showed that both FFTW3 and iMKL performed significantly better when launched with 2 MPI tasks/node, which corresponds to 1 NUMA domain/MPI task. S3DFT was found to perform similarly in the 1 MPI task/node and 2 MPI tasks/node configurations. Therefore, here we present results obtained using 2 MPI tasks/node for FFTW3 and iMKL. For S3DFT, in the small problem scale, we used the 1 MPI task/node configuration, and in the large problem scale, the 2 MPI tasks/node configuration.

\begin{figure}[h!]
    \centering
    \begin{subfigure}{0.49\textwidth}
        \includegraphics[scale=0.165]{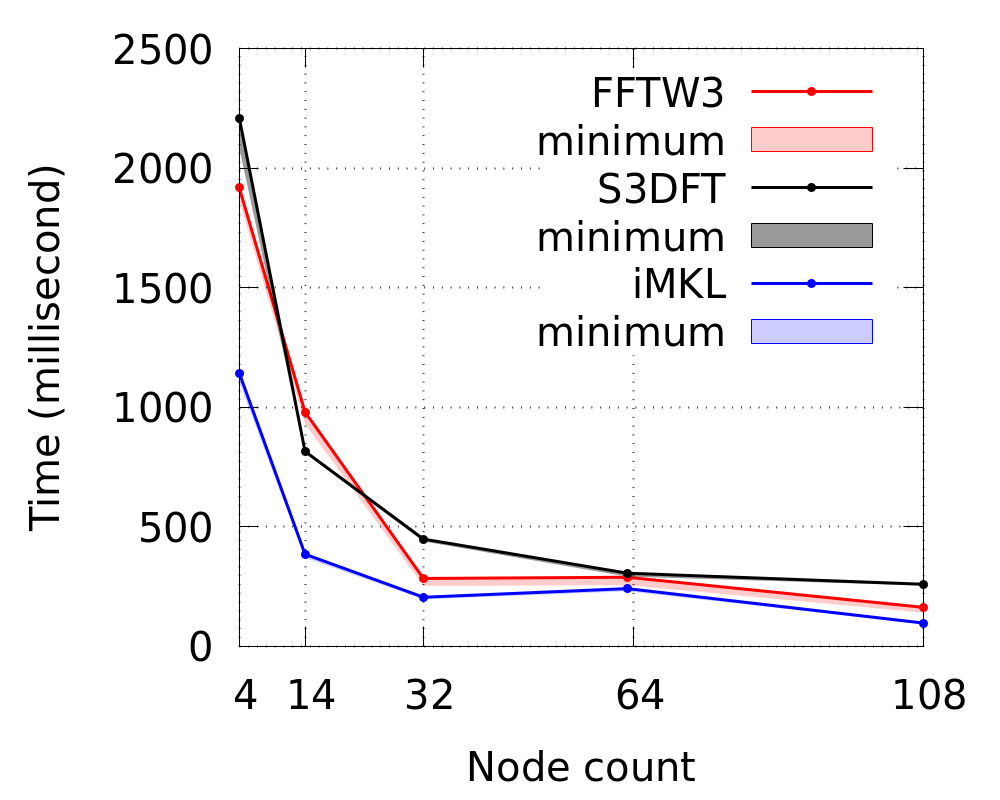}
    \end{subfigure}
    \begin{subfigure}{0.49\textwidth}
        \includegraphics[scale=0.165]{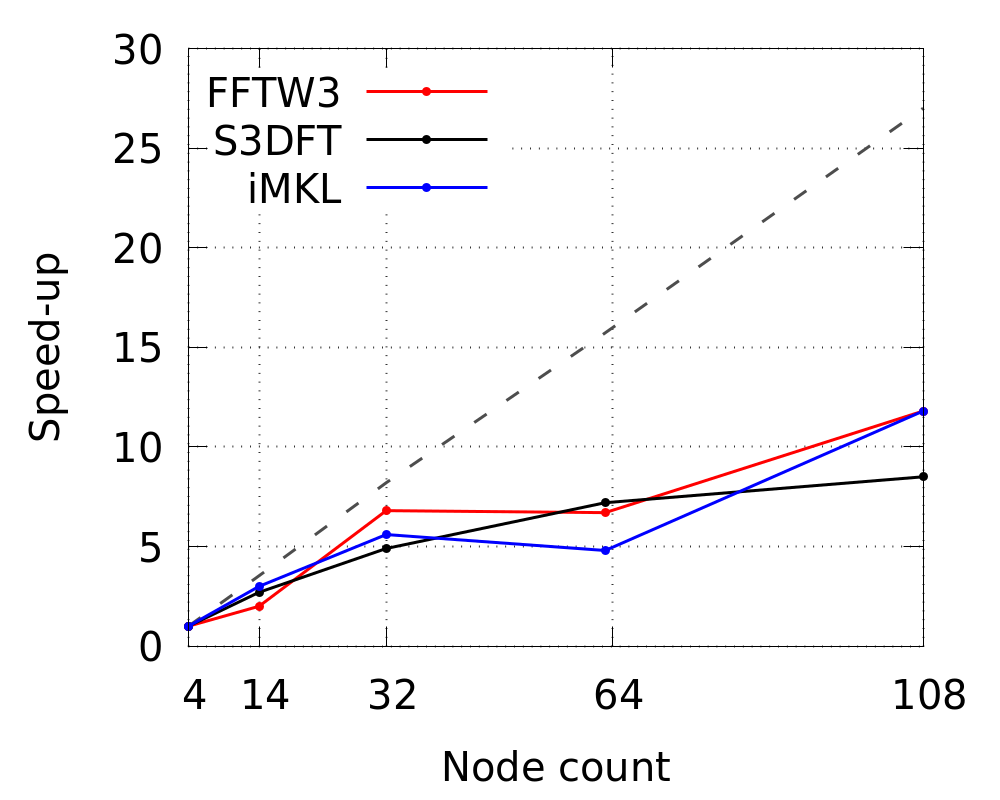}
    \end{subfigure}
    \caption{Strong scaling performance comparison of S3DFT, FFTW3 and iMKL for problem size $N=840$. Left and right panels show the time-to-solution and speed-up, respectively. The grey dashed line indicates the ideal linear scaling.}
    \label{fig:api_strong_scaling_small}
\end{figure}

\begin{figure}[h!]
    \centering
    \begin{subfigure}{0.49\textwidth}
        \includegraphics[scale=0.165]{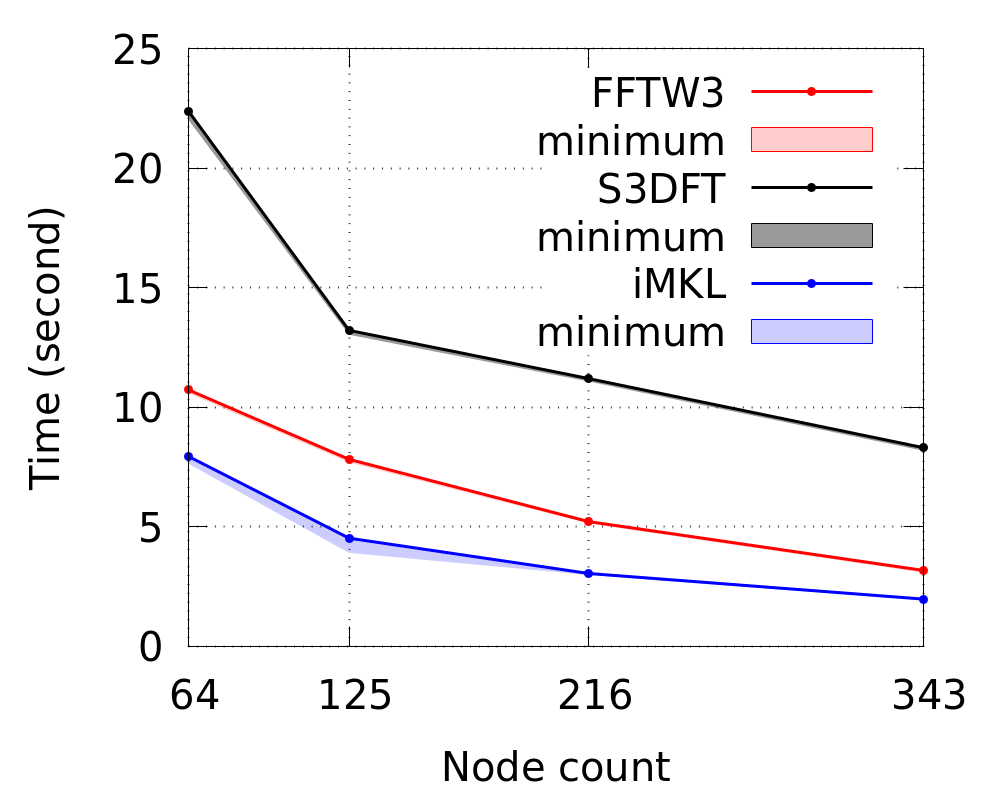}
    \end{subfigure}
    \begin{subfigure}{0.49\textwidth}
        \includegraphics[scale=0.165]{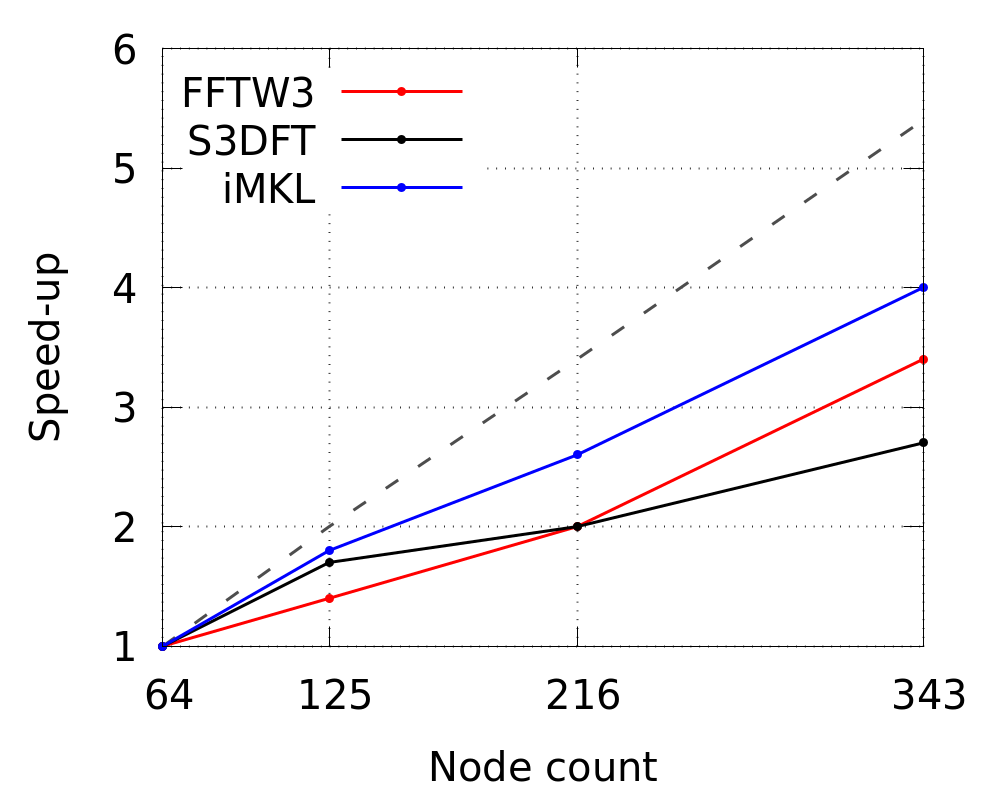}
    \end{subfigure}
    \caption{Strong scaling performance comparison of S3DFT, FFTW3 and iMKL for problem size $N=3360$. Left and right panels show the time-to-solution and speed-up, respectively. The grey dashed line indicates the ideal linear scaling.}
    \label{fig:api_strong_scaling_large}
\end{figure}

In what we call small problem scale, strong scaling comparisons were conducted for problem sizes $N=120,240,480,600,840$.  Similarly, in the large problem scale, we ran strong scaling tests for problem sizes $N=2520,3360,4200$. The results for sizes $N=840,3360$ are provided in figures~\ref{fig:api_strong_scaling_small} and~\ref{fig:api_strong_scaling_large}. In both the small and large problem scales, we observed similar scaling behaviours and performances for all investigated problem sizes.


Results show that iMKL was consistently the fastest across all problem sizes and node counts, followed by FFTW3. In the small problem scale, on an average, iMKL was 1.8 times faster than S3DFT, while FFTW3 was found to be 1.2 times faster than S3DFT. Here, we also observed that S3DFT was frequently slightly faster than FFTW3 for node counts $<32$. In the large problem scale, on an average, iMKL was 3.2 times faster than S3DFT, and FFTW3 was 2.0 faster than S3DFT. Interestingly, for large sizes and node counts, S3DFT was found to scale less efficiently than its competitors, which can be attributed to the poor scaling of algorithms~\ref{alg:adapted_cannonalgo_rho_l} and~\ref{alg:adapted_cannonalgo_transposed_rho_r}, as discussed in subsection~\ref{subsec:dm_tmm}.

%%%%%%%%%%%%%%%%%%%%%%%%%%%%%%%%%%%%%%%%%%%%
\section{Conclusion}
\label{sec:conclusion}
We have presented a new parallel algorithm called 3D-DFT by BTMM that exploits block tensor-matrix multiplication to compute the 3D-DFT of a cubic domain using point-to-point communication. The algorithm is implemented as a C++ library called S3DFT capable of utilizing shared memory parallelism across multiple NUMA domains within a single compute node. In the process, we designed, developed and tested three adapted variants of Cannon's algorithm. These adaptations enable the use of tensor operands, realize multi-level parallelism, make efficient use of the technique of overlapping computation and communication with the help of a custom work-sharing function for OpenMP threads, and eliminate additional communication overheads of a combined transpose-multiply-transpose operation by remapping the mesh of PEs instead of transposing the data. Its implementation has been optimized for the JUWELS Cluster, and its core functions analyzed to show its efficiency, and acknowledge its shortcomings. The performance of S3DFT was compared with those of competitive, well-known libraries for a wide range of problem sizes.

Our analysis by micro-benchmarking has shown that our shared memory tensor-matrix multiplication reaches $88\%$ of the single node peak performance. Of the three variants of the distributed memory tensor-matrix multiplication algorithms, one scales excellently while the others scale poorly. We identified the origin of this behaviour in their intrinsic communication patterns.  This is the main cause for the observed poor scaling performance of S3DFT when compared to FFTW3 and iMKL. Further efforts to improve S3DFT should therefore focus on the scalability and performance of algorithms~\ref{alg:adapted_cannonalgo_rho_l} and~\ref{alg:adapted_cannonalgo_transposed_rho_r}.

At the current stage, the 3D-DFT by BTMM algorithm is not a viable alternative to modern FFT-based approaches on computer clusters with fast node interconnects. Different results might be expected on different computer clusters. For example, it is very possible that S3DFT shows superior performance compared to FFTW3 and iMKL when run on a computer cluster using a network with higher latency than the Mellanox InfiniBand network of the JUWELS Cluster. This is an interesting question that could inspire further investigations.

%++++++++++++++++++++++++++++++++++++++++++++
% special: acknowledgements
%++++++++++++++++++++++++++++++++++++++++++++

%\subsubsection{Acknowledgements}
PC and DM acknowledge funding from the Helmholtz European Partnership program "Innovative high-performance computing approaches for molecular neuro-medicine". PC acknowledges funding from the Human Brain Project (EU Horizon 2020). This research was supported by the Joint Lab ``Supercomputing and Modeling for the Human Brain''. NM thanks the support team at JSC, especially Ilya Zhukov, and Rolf Rabenseifner for all the helpful discussions and suggestions.

%++++++++++++++++++++++++++++++++++++++++++++
% bibliography
%++++++++++++++++++++++++++++++++++++++++++++
\bibliographystyle{splncs04}
\bibliography{references}

\begin{thebibliography}{10}
\providecommand{\url}[1]{\texttt{#1}}
\providecommand{\urlprefix}{URL }
\providecommand{\doi}[1]{https://doi.org/#1}

\bibitem{alvarez2021}
Alvarez, D.: {JUWELS} {C}luster and {B}ooster: {E}xascale {P}athfinder with
  {M}odular {S}upercomputing {A}rchitecture at {J}uelich {S}upercomputing
  {C}entre. Journal of large-scale research facilities JLSRF  \textbf{7} (10
  2021). \doi{10.17815/jlsrf-7-183}

\bibitem{arnold2013}
Arnold, A., Fahrenberger, F., Holm, C., Lenz, O., Bolten, M., Dachsel, H.,
  Halver, R., Kabadshow, I., G\"ahler, F., Heber, F., Iseringhausen, J.,
  Hofmann, M., Pippig, M., Potts, D., Sutmann, G.: Comparison of scalable fast
  methods for long-range interactions. Phys. Rev. E  \textbf{88},  063308 (Dec
  2013). \doi{10.1103/PhysRevE.88.063308},
  \url{https://link.aps.org/doi/10.1103/PhysRevE.88.063308}

\bibitem{ayala2021}
Ayala, A., Tomov, S., Stoyanov, M., Dongarra, J.: Scalability issues in fft
  computation. In: Malyshkin, V. (ed.) Parallel Computing Technologies. pp.
  279--287. Springer International Publishing, Cham (2021)

\bibitem{campa2009}
Campa, A., Dauxois, T., Ruffo, S.: Statistical mechanics and dynamics of
  solvable models with long-range interactions. Physics Reports
  \textbf{480}(3),  57--159 (2009).
  \doi{https://doi.org/10.1016/j.physrep.2009.07.001},
  \url{https://www.sciencedirect.com/science/article/pii/S0370157309001586}

\bibitem{cannon1969}
Cannon, L.E.: A Cellular Computer to Implement the Kalman Filter Algorithm.
  Ph.D. thesis, USA (1969), aAI7010025

\bibitem{intel2021}
Corporation, I.: Where can i find information about flops per cycle for
  intel(r) processors?
  \url{https://www.intel.com/content/www/us/en/support/articles/000057415/processors.html}
  (2021), last accessed: 08.11.2022

\bibitem{frigo2005}
Frigo, M., Johnson, S.: The design and implementation of fftw3. Proceedings of
  the IEEE  \textbf{93}(2),  216--231 (2005). \doi{10.1109/JPROC.2004.840301}

\bibitem{fftw3_manual}
Frigo, M., Johnson, G.S.: FFTW. Massachusetts Institute of Technology (December
  2020), available at \url{http://www.fftw.org/fftw3.pdf}

\bibitem{gupta1993}
Gupta, A., Kumar, V.: Scalability of parallel algorithms for matrix
  multiplication. In: 1993 International Conference on Parallel Processing -
  ICPP'93. vol.~3, pp. 115--123 (1993). \doi{10.1109/ICPP.1993.160}

\bibitem{hager2010}
Hager, G., Wellein, G.: Introduction to High Peformance Computing for
  Scientists and Engineers (07 2010). \doi{10.1201/EBK1439811924}

\bibitem{harvey2009}
Harvey, M.J., De~Fabritiis, G.: An implementation of the smooth particle mesh
  ewald method on gpu hardware. Journal of Chemical Theory and Computation
  \textbf{5}(9),  2371--2377 (2009). \doi{10.1021/ct900275y},
  \url{https://doi.org/10.1021/ct900275y}, pMID: 26616618

\bibitem{intel2020}
Intel Corporation: Intel(R) Xeon(R) Processor Scalable Family Specification
  Update, 017 edn. (October 2020), available at
  \url{https://www.intel.com/content/dam/www/public/us/en/documents/specification-updates/xeon-scalable-spec-update.pdf}

\bibitem{intel2022}
Intel Corporation: Intel(R) Advisor User Guide, 2022.3 edn. (2022), available
  at
  \url{https://www.intel.com/content/www/us/en/develop/documentation/advisor-user-guide/top.html}

\bibitem{jung2016}
Jung, J., Kobayashi, C., Imamura, T., Sugita, Y.: Parallel implementation of 3d
  fft with volumetric decomposition schemes for efficient molecular dynamics
  simulations. Computer Physics Communications  \textbf{200},  57--65 (2016).
  \doi{https://doi.org/10.1016/j.cpc.2015.10.024},
  \url{https://www.sciencedirect.com/science/article/pii/S0010465515004063}

\bibitem{kloeffel2020}
Klöffel, T., Mathias, G., Meyer, B.: Integrating state of the art compute,
  communication, and autotuning strategies to multiply the performance of the
  application programm cpmd for ab initio molecular dynamics simulations
  (2020). \doi{10.48550/ARXIV.2003.08477},
  \url{https://arxiv.org/abs/2003.08477}

\bibitem{kohnke2020}
Kohnke, B., Kutzner, C., Grubmüller, H.: A gpu-accelerated fast multipole
  method for gromacs: Performance and accuracy. Journal of Chemical Theory and
  Computation  \textbf{16}(11),  6938--6949 (2020).
  \doi{10.1021/acs.jctc.0c00744},
  \url{https://doi.org/10.1021/acs.jctc.0c00744}, pMID: 33084336

\bibitem{kolda2009}
Kolda, T., Bader, B.: Tensor decompositions and applications. SIAM Review
  \textbf{51},  455--500 (08 2009). \doi{10.1137/07070111X}

\bibitem{lippert1997}
Lippert, T., Schilling, K., Trentmann, S., Toschi, F., Tripiccione, R.: Fft for
  the ape parallel computer. International Journal of Modern Physics C
  \textbf{8}(06),  1317--1334 (1997)

\bibitem{s3dft}
Malapally, N.: S3dft: Scalable 3d-dft. Available on
  \url{https://gitlab.com/anxiousprogrammer/s3dft} (2021), last accessed:
  08.11.2022

\bibitem{tixl}
Malapally, N.: Tixl: Timed experiments in a loop. Available on
  \url{https://gitlab.com/anxiousprogrammer/tixl} (2021), last accessed:
  08.11.2022

\bibitem{pekurovsky2011}
Pekurovsky, D.: Ultrascalable fourier transfroms in three dimensions. In:
  Proceedings of the 2011 TeraGrid Conference: Extreme Digital Discovery. TG
  '11, Association for Computing Machinery, New York, NY, USA (2011).
  \doi{10.1145/2016741.2016751}, \url{https://doi.org/10.1145/2016741.2016751}

\bibitem{quintin2013}
Quintin, J.N., Hasanov, K., Lastovetsky, A.: Hierarchical parallel matrix
  multiplication on large-scale distributed memory platforms. In: 2013 42nd
  International Conference on Parallel Processing. pp. 754--762 (2013).
  \doi{10.1109/ICPP.2013.89}

\bibitem{rockmore2000}
Rockmore, D.: The fft: an algorithm the whole family can use. Computing in
  Science Engineering  \textbf{2}(1),  60--64 (2000). \doi{10.1109/5992.814659}

\bibitem{sedukhin2015}
Sedukhin, S., Sakai, T., Nakasato, N.: 3d discrete transforms with cubical data
  decomposition on the ibm blue gene/q. Proceedings of the 30th International
  Conference on Computers and Their Applications, CATA 2015 pp. 193--200 (01
  2015)

\bibitem{toukmaji1996}
Toukmaji, A.Y., Board, J.A.: Ewald summation techniques in perspective: a
  survey. Computer Physics Communications  \textbf{95}(2),  73--92 (1996).
  \doi{https://doi.org/10.1016/0010-4655(96)00016-1},
  \url{https://www.sciencedirect.com/science/article/pii/0010465596000161}

\bibitem{tuckerman2010}
Tuckerman, M.: Statistical Mechanics: Theory and Molecular Simulation. Oxford
  Graduate Texts, OUP Oxford (2010),
  \url{https://books.google.de/books?id=Lo3Jqc0pgrcC}

\bibitem{yildirim2015}
Yildirim, H., Matos, J., Kara, A.: Role of long-range interactions for the
  structure and energetics of olympicene radical adsorbed on au(111) and
  pt(111) surfaces. The Journal of Physical Chemistry C  \textbf{119}(45),
  25408--25419 (2015). \doi{10.1021/acs.jpcc.5b08191},
  \url{https://doi.org/10.1021/acs.jpcc.5b08191}

\end{thebibliography}
\end{document}